\documentclass[reprint,superscriptaddress,nofootinbib,amsmath,amssymb,aps]{revtex4-1}

\usepackage{graphicx}
\usepackage{dcolumn}
\usepackage{bm}
\usepackage[export]{adjustbox}
\usepackage{xcolor}

\newcommand{\kms}{\ensuremath{\mathrm{km\,s}^{-1}}}

\def\mnras{Mon. Not. R. Astron. Soc.}

\def\aap{Astron. Astrophys.}

\begin{document}

\title{Stellar rotation of S301 as a macroscopic gyroscope to test general relativity}

\author{Pau Amaro Seoane}
\affiliation{Universitat Polit\`ecnica de Val\`encia, Spain}
\affiliation{Max Planck Institute for Extraterrestrial Physics, Garching, Germany}
\affiliation{Kavli Institute for Astronomy and Astrophysics at Peking University, Beijing, China}

\author{Xian Chen}
\affiliation{Kavli Institute for Astronomy and Astrophysics at Peking University, Beijing, China}
\affiliation{Astronomy Department, School of Physics, Peking University, Beijing, China}

\author{Alejandro Torres-Orjuela}
\email{Corresponding author: atorreso@bimsa.cn}
\affiliation{Beijing Institute of Mathematical Sciences and Applications, Beijing, China}

\author{Reinhard Genzel}
\affiliation{Max Planck Institute for Extraterrestrial Physics, Giessenbachstra{\ss}e 1, 85748 Garching, Germany}
\affiliation{Departments of Physics \& Astronomy, Le Conte Hall, University of California, Berkeley, CA 94720, USA}

\author{Frank Eisenhauer}
\affiliation{Max Planck Institute for Extraterrestrial Physics, Giessenbachstra{\ss}e 1, 85748 Garching, Germany}
\affiliation{Department of Physics, TUM School of Natural Sciences, Technical University of Munich, 85748 Garching, Germany}

\author{Thomas Ott}
\affiliation{Max Planck Institute for Extraterrestrial Physics, Giessenbachstra{\ss}e 1, 85748 Garching, Germany}

\author{Stefan Gillessen}
\affiliation{Max Planck Institute for Extraterrestrial Physics, Giessenbachstra{\ss}e 1, 85748 Garching, Germany}

\author{Guillaume Bourdarot}
\affiliation{Max Planck Institute for Extraterrestrial Physics, Giessenbachstra{\ss}e 1, 85748 Garching, Germany}

\author{Diogo C. Ribeiro}
\affiliation{Max Planck Institute for Extraterrestrial Physics, Giessenbachstra{\ss}e 1, 85748 Garching, Germany}

\author{Matteo Sadun Bordoni}
\affiliation{Max Planck Institute for Extraterrestrial Physics, Giessenbachstra{\ss}e 1, 85748 Garching, Germany}

\author{Simran Joharle}
\affiliation{Max Planck Institute for Extraterrestrial Physics, Giessenbachstra{\ss}e 1, 85748 Garching, Germany}

\author{Felix Mang}
\affiliation{Max Planck Institute for Extraterrestrial Physics, Giessenbachstra{\ss}e 1, 85748 Garching, Germany}
\affiliation{Department of Physics, TUM School of Natural Sciences, Technical University of Munich, 85748 Garching, Germany}

\author{Andreas Burkert}
\affiliation{University Observatory, Faculty of Physics, Ludwig-Maximilians-Universit{\"a}t, Scheinerstra{\ss}e 1, 81679 Munich, Germany}
\affiliation{Max Planck Institute for Extraterrestrial Physics, Giessenbachstra{\ss}e 1, 85748 Garching, Germany}

\author{Jorge Cuadra}
\affiliation{Universidad Adolfo Ib{\'a}\~nez, Av. Padre Hurtado 750, Vi\~na del Mar, Chile}
\affiliation{Millennium Nucleus on Transversal Research and Technology to Explore Supermassive Black Holes (TITANS), Chile}

\author{Diego Calderón}
\affiliation{Max-Planck-Institut für Astrophysik, Karl-Schwarzschild-Straße 1, 85748 Garching, Germany}

\author{Hagai B. Perets}
\affiliation{Physics department, Technion - Israel Institute of Technology, Technion city, Haifa 3200002, Israel}

\author{Tsvi Piran}
\affiliation{Racah Institute of Physics, The Hebrew University, Jerusalem 91904, Israel}

\author{Thorsten Naab}
\affiliation{Max Planck Institute for Astrophysics, Karl-Schwarzschild-Stra{\ss}e 1, 85748 Garching, Germany}

\author{Re'em Sari}
\affiliation{Racah Institute of Physics, The Hebrew University, Jerusalem 91904, Israel}

\date{\today}

\begin{abstract}
Stellar trajectories around the Galactic Center provide a testing environment for general relativity. The intrinsic rotation of these stars evolves under covariant transport in curved spacetime and classical Newtonian quadrupole torques. We analyze the recently observed S301 S-star to quantify the relativistic precession of its rotational axis. Its 8.7-year period and eccentricity of $e = 0.982$ localize geodetic precession and Newtonian quadrupole torques to a step function at periapsis. We incorporate first-order post-Newtonian corrections into the orbital kinematics to calculate the spatial trajectory. Sampling an isotropic distribution of initial orientations and viewing geometries over a 40-year period across a grid of equatorial velocities and rotational ellipticities, we calculate the statistical likelihood of an absolute shift in the projected rotational line broadening, $|\Delta v \sin i|$. The relativistic geodetic shift scales linearly with $v_{\rm rot}$ and the classical quadrupole shift is independent of rotation speed, scaling with $q$. The absolute maximum velocity shift saturates at $46.1\,\kms$ for oblate stars. The absolute median shifts, driven by geodetic precession, range from $3\,\kms$ to $6.3\,\kms$. We calculate the time-domain observable $|\Delta v \sin i|$ to provide a target for infrared spectrographs testing the Schwarzschild metric around Sgr~A$^\ast$. The spin of S301 acts as a flying gyroscope whose drift, if measured, can test Einstein's theory in a regime that has not previously been accessible.
\end{abstract}

\maketitle

\section{Introduction}

The Galactic Center contains Sgr~A$^\ast$, a black hole with a mass of $M = 4.3 \times 10^6\,M_\odot$. The S-stars orbiting Sgr~A$^\ast$ provide a testing environment for general relativity. Prior studies measured the first-order post-Newtonian Schwarzschild precession and the combined transverse Doppler and gravitational redshift by tracking the orbit of the star S2 \cite{GRAVITY2020, GRAVITY2018, DoEtAl2019}.

A rotating body in free fall parallel-transports its intrinsic rotational axis according to Fermi-Walker transport. This covariant transport induces prograde geodetic precession. The orientation of the rotational axis alters the inclination angle $i$ between the stellar pole and the line of sight of the observer. Spectroscopic absorption line broadening measures this time-varying projected rotational velocity, $v \sin i$ \cite{DaviesEtAl2021}.

A rotating star exhibits an equatorial bulge, generating a mass quadrupole moment. As the star moves through the spatial gradient of the gravitational potential, this oblateness generates a Newtonian torque that precesses the rotational axis. For extended sources, this classical rigid-body interaction overlaps with the relativistic signal.

The GRAVITY+ collaboration reported S301, a main-sequence star with an apparent magnitude of $m_K = 19.3$ \cite{S3012026}. The star possesses an 8.7-year period and reaches a velocity of $v_p = 25000\,\kms$ at periapsis, resulting in an eccentricity of $e = 0.982$. The orbital parameters of S301 isolate the relativistic signal from the Newtonian effects. The eccentricity localizes the precessional evolution into a discrete change at periapsis, and the physical radius of S301 limits the quadrupole torque relative to the covariant geodetic precession.

Because observations do not currently constrain the rotational velocity and oblateness of S301, a systematic analysis of its physical parameter space is necessary. We calculate the relativistic precession of the rotational axis of S301 over a 40-year baseline across a grid of rotational velocities and ellipticities. Integrating the covariant transport and first-order post-Newtonian orbital kinematics demonstrates that the projected rotational velocity $v \sin i$ undergoes a step-function variation during periapsis passages. The statistical analysis of an isotropic orientation distribution indicates absolute median velocity shifts ranging from $3\,\kms$ to $6.3\,\kms$, with absolute maximum shifts approaching $46.1\,\kms$ for specific geometric configurations. Tracking these velocity steps over multiple pericenter passages utilizes the post-Newtonian orbital advance to break the degeneracy between the relativistic precession and the classical quadrupole torque. We demonstrate how the physical properties of S301 modulate the Newtonian quadrupole torque, mapping the threhsolds required for detection. Because the mass of Sgr~A$^\ast$ is well-constrained by orbital astrometry, isolating the geodetic precession provides an independent test of the covariant transport of spin, directly constraining post-Newtonian deviations in the spatial curvature of the metric.

\section{Orbital kinematics and covariant dynamics}

To calculate the orbital trajectory of S301, we apply the first-order post-Newtonian equations of motion for a test particle in the Schwarzschild metric. Using geometric units ($G=c=1$), the acceleration $\bm{a}$ is

\begin{equation}
\label{eq:1pn}
\bm{a} = -\frac{M}{r^3}\bm{r} + \frac{M}{r^3} \left[ \left( \frac{4M}{r} - v^2 \right) \bm{r} + 4 (\bm{r} \cdot \bm{v}) \bm{v} \right],
\end{equation}

\noindent
where the first term represents Newtonian gravity and the bracketed term accounts for relativistic orbital precession.

The Mathisson-Papapetrou-Dixon equations govern the covariant transport of the stellar intrinsic angular momentum tensor $S^{\mu\nu}$. By defining the proper time $\tau$, the four-velocity $u^\mu$, and the four-momentum $p^\mu$, the transport along the worldline obeys

\begin{equation}
\label{eq:mpd}
\frac{D S^{\mu\nu}}{d\tau} = p^\mu u^\nu - p^\nu u^\mu + \mathcal{T}^{\mu\nu},
\end{equation}

\noindent
where the covariant derivative $D/d\tau$ projects the geometry of the spacetime, and $\mathcal{T}^{\mu\nu}$ represents the torque generated by the coupling of higher-order mass multipoles to the Riemann curvature tensor. We impose the Tulczyjew-Dixon spin-supplementary condition, $S^{\mu\nu} p_\nu = 0$, to specify the center of mass of the extended body.

In the weak-field and slow-motion limit, assuming geodesic motion for the center of mass, the evolution of the spatial intrinsic angular momentum vector $\bm{S}$ separates into relativistic and classical components, producing

\begin{equation}
\label{eq:spin_evolution_S}
\frac{d\bm{S}}{dt} = \bm{\Omega}_{\text{geod}} \times \bm{S} + \bm{N}_{\text{quad}},
\end{equation}

\noindent
where the mass $M$ of the black hole determines the geodetic precession $\bm{\Omega}_{\text{geod}}$, and the local tidal field exerts the Newtonian torque $\bm{N}_{\text{quad}}$. We omit the Lense-Thirring precession because the rotation of the black hole introduces a negligible correction compared to the geodetic term, reducing the system to the Schwarzschild geometry.

Using the orbital distance $r$, the orbital velocity $\bm{v}$, and the unit direction vector $\bm{n} = \bm{r}/r$, the geodetic precession vector evaluates to

\begin{equation}
\label{eq:geod_prec}
\bm{\Omega}_{\text{geod}} = \frac{3M}{2r^2} (\bm{n} \times \bm{v}).
\end{equation}

\noindent
The magnitude of the intrinsic angular momentum is $S = I_3 \omega_{\text{rot}}$, where $\omega_{\text{rot}}$ represents the angular rotation velocity, and $I_3$ is the principal moment of inertia along the rotational axis. Defining the rotational ellipticity as $q = (I_3 - I_1)/I_3$, we divide Eq.~(\ref{eq:spin_evolution_S}) by $S$ to obtain the evolution of the unit rotation vector $\bm{s} = \bm{S}/S$, resulting in

\begin{equation}
\label{eq:spin_evolution_s}
\frac{d\bm{s}}{dt} = \left(\bm{\Omega}_{\text{geod}} - \bm{\Omega}_{\text{quad}}\right) \times \bm{s}.
\end{equation}

\noindent
As detailed in the Appendix, evaluating the classical rigid-body torque produces the Newtonian quadrupole precession vector

\begin{equation}
\label{eq:quad_prec}
\bm{\Omega}_{\text{quad}} = 3 \frac{M}{r^3} \frac{q}{\omega_{\text{rot}}} (\bm{n} \cdot \bm{s}) \bm{n}.
\end{equation}

\section{Periapsis scaling and the S301 configuration}

Interferometric observations constrain the orbit of S301 and provide an eccentricity of $e = 0.982$ \cite{S3012026}. This parameter isolates the relativistic scaling factors governing the evolution of the rotational axis.

We calculate the peak precession rates by evaluating the orbital kinematics at periapsis, where the distance is $r_p = a(1-e)$. The velocity vector $\bm{v}$ aligns perpendicular to $\bm{n}$, resulting in a peak velocity $v_p = [M(1+e)/(a(1-e))]^{1/2}$. Substituting these values into Eq.~(\ref{eq:geod_prec}) calculates the peak geodtic precession rate

\begin{equation}
\label{eq:geod_peak}
|\bm{\Omega}_{\text{geod,p}}| = \frac{3}{2} M^{3/2} a^{-5/2} (1+e)^{1/2} (1-e)^{-5/2}.
\end{equation}

\noindent
Applying the periapsis distance to Eq.~(\ref{eq:quad_prec}) defines the maximum quadrupole precession scaling relation

\begin{equation}
\label{eq:quad_peak}
\max|\bm{\Omega}_{\text{quad,p}}| = 3 M q \omega_{\text{rot}}^{-1} a^{-3} (1-e)^{-3}.
\end{equation}

\noindent
These relations show that the $1-e$ scaling factors confine the precessional evolution to a short duration near periapsis. For S301, the variation acts as a discrete step function.

At an apparent magnitude of $m_K = 19.3$, S301 restricts the magnitude of the Newtonian signal. Modeling S301 as a main-sequence star sets a radius of $R_{\ast} = 1.5\,R_\odot$. The angular velocity $\omega_{\text{rot}}$ equals $v_{\text{rot}}/R_{\ast}$, which makes $\bm{\Omega}_{\text{quad}}$ linearly proportional to $R_{\ast}$. The physical radius of S301 limits the Newtonian precession relative to the geodetic signal.

\section{Step-function observables and parametric scaling}

We quantify the dependence of the projected rotational velocity shift on the viewing geometry and the intrinsic stellar parameters using a Monte Carlo method. We integrate the covariant transport over a 40-year observation baseline. The numerical integration models the trajectory using Newtonian orbital mechanics augmented by first-order post-Newtonian corrections. We incorporate the Einstein-Infeld-Hoffmann acceleration to account for relativistic Schwarzschild periapsis precession. This precession rotates the orbital ellipse relative to the line of sight. Because the observable $\Delta v \sin i$ isolates the projection of the rotational axis along this fixed line of sight, the orbital advance alters the viewing geometry during successive periapsis passages, modulating the projected rotational velocity shift.

We compute the transport through Eq.~(\ref{eq:spin_evolution_s}). To evaluate the impact of stellar structure and kinematics on the measurable signal, the simulation explores a parameter grid specifying equatorial velocities $v_{\text{rot}} \in \{200, 300, 350, 400\}\,\kms$ and rotational ellipticities $q \in \{0.05, 0.1, 0.3\}$ \cite{Barker2020,BolmontEtAl2016,XuEtAl2025}. For each of the 12 parameter combinations, we compute 2000 realizations by sampling an isotropic distribution of initial orientations and observer lines of sight to randomize the geometric configuration.

The variation operates as a discrete step localized at the periapsis passage. The amplitude of the projected velocity shift correlates with the geometric configuration of the rotational axis relative to the orbital plane and the observer. The geodetic torque precesses the rotational axis around the orbital angular momentum vector. Maximum displacements occur when the initial rotational axis lies near the orbital plane, the orbit is viewed near edge-on, and the observer views the star near pole-on.

The data demonstrate physical scaling laws for the two driving torques. The equation $\Delta v \sin i \approx v_{\text{rot}} \cos i \Delta i$ approximates the accumulated shift in line-of-sight velocity, where $\Delta i$ is the angular displacement of the rotational axis relative to the observer. Because the geodetic precession rate $\bm{\Omega}_{\text{geod}}$ acts independently of the stellar properties, its resulting angular displacement $\Delta i_{\text{geod}}$ remains constant for a given orbit. The geodetic contribution to the velocity shift scales linearly with $v_{\text{rot}}$.

The Newtonian quadrupole precession rate depends on the rotation parameters. Equation~(\ref{eq:quad_prec}) indicates that the quadrupole precession rate scales inversely with the rotation rate, $\bm{\Omega}_{\text{quad}} \propto q/v_{\text{rot}}$. Integrating this rate to find the angular displacement results in $\Delta i_{\text{quad}} \propto q/v_{\text{rot}}$. When projecting this back into the observable velocity space, the $v_{\text{rot}}$ dependencies cancel, leaving $\Delta v_{\text{quad}} \approx v_{\text{rot}} \Delta i_{\text{quad}} \propto q$. The observable velocity shift arising from the classical quadrupole torque scales solely with the ellipticity $q$ and stellar radius, independent of the rotational velocity.

\begin{figure}[t]
    \centering
    \includegraphics[width=\columnwidth]{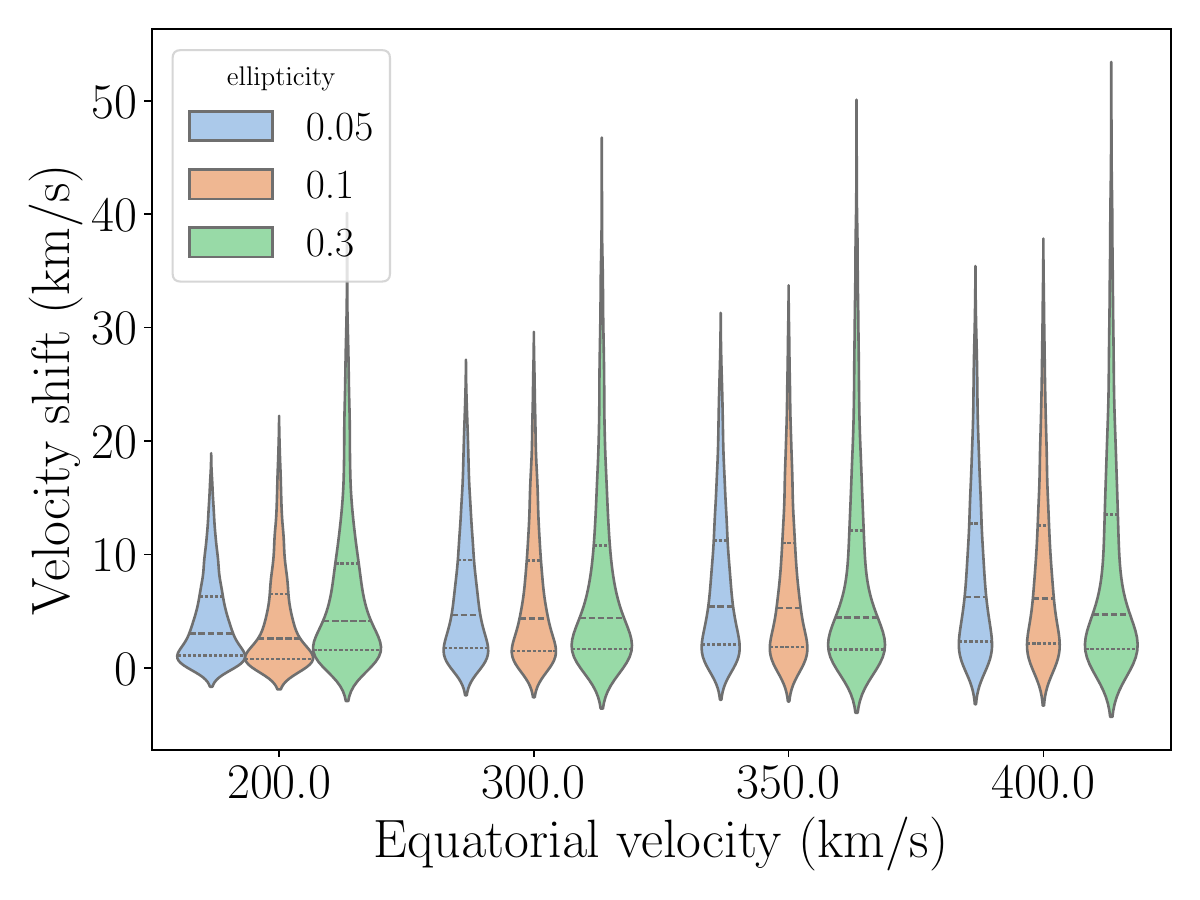}
    \caption{Statistical distribution of the absolute projected rotational velocity shift, $|\Delta v \sin i|$, for S301 accumulated over 40 years across the parameter grid of equatorial velocities, $v_{\text{rot}}$, and rotational ellipticities, $q$. The distributions widen at higher ellipticities due to the introduction of Newtonian quadrupole torques. We note that the visual extension of the distributions into negative values is an artifact of the kernel density estimation smoothing; the underlying data are strictly non-negative.}
    \label{fig:distributions}
\end{figure}

\begin{figure}[t]
    \centering
    \includegraphics[width=\columnwidth]{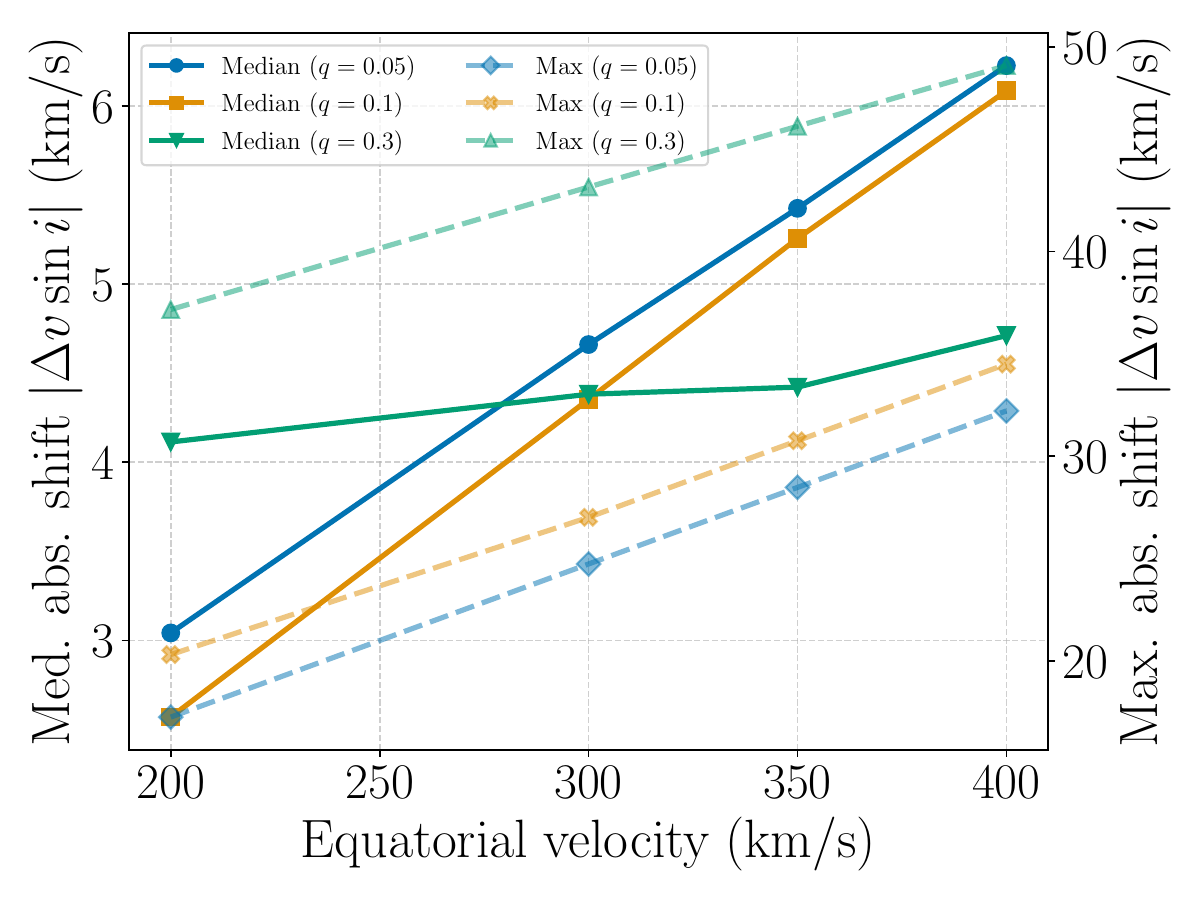}
    \caption{Absolute median (solid lines) and maximum $|\Delta v \sin i|$ shifts (dashed lines) across 2000 sampled isotropic orientations after 40 years. The absolute median shift demonstrates a linear growth with $v_{\text{rot}}$. The absolute maximum shift reaches a saturation limit for oblate stars where the quadrupole torque establishes a rotation-independent baseline.}
    \label{fig:combined}
\end{figure}

The numerical integrations presented in Figures \ref{fig:distributions} and \ref{fig:combined} display the absolute value of the accumulated projected velocity shift, $|\Delta v \sin i|$, over the full 40-year baseline. This duration covers between four and five pericenter passages. Over long timescales, the physical precession drives the rotational axis to complete full precessional cycles. The inclination angle and the measured velocity oscillate within the geometric bounds of the actual stellar rotational velocity, rather than growing monotonically. Tracking the absolute magnitude over the initial 40-year window quantifies the signal amplitude generated across multiple steps.

For a star with $q=0.05$, the geodetic term dominates, and the absolute maximum observed velocity shift increases from $16.9\,\kms$ at $v_{\text{rot}}=200\,\kms$ to $32.1\,\kms$ at $v_{\text{rot}}=400\,\kms$. A star with $q=0.3$ introduces quadrupole shifts reaching $32.5\,\kms$ at $v_{\text{rot}}=200\,\kms$. Because the quadrupole velocity shift acts independently of rotation, increasing the rotation speed to $400\,\kms$ adds the linear geodetic baseline, resulting in a maximum absolute shift of $46.1\,\kms$.

The absolute median shifts across the isotropic distribution remain bound between $3\,\kms$ and $6.3\,\kms$. Current near-infrared spectrographs operate with a velocity resolution limit of $50\,\kms$. To produce a shift exceeding this threshold, the physical parameters of S301 must occupy the upper bounds of their modeled intervals, $q=0.3$ and $v_{\text{rot}}=400\,\kms$, alongside an edge-on viewing geometry. The median shifts match the planned capabilities of next-generation large telescopes and infrared spectrographs, which target velocity resolutions of $3\,\kms$.

Measuring the step-like changes across successive pericenter passages separates the geodetic precession from the classical tidal torque. The relativistic orbital advance alters the orientation of the orbital ellipse relative to the observer line of sight. This geometric variation modulates the amplitude of the observed velocity shift over time. Because the relativistic precession depends on the mass of the central black hole and the stellar rotation speed, while the classical tidal torque depends on the stellar ellipticity, tracking the modulated signal across sequential orbits breaks the degeneracy between the two effects and isolates the relativistic component.

\section{Conclusions}

The intrinsic rotation of stars on bound orbits around Sgr~A$^\ast$ provides a dynamical test of general relativity in the strong-field regime. Newtonian quadrupole torques arising from stellar oblateness introduce a degeneracy with the relativistic geodetic signal. The orbital geometry of the recently observed S301 S-star, characterized by an 8.7-year period and an eccentricity $e=0.982$ \cite{S3012026}, suppresses these classical interactions by confining their action to a narrow temporal window around periapsis, while the stellar radius limits their absolute magnitude. Evaluating the Mathisson-Papapetrou-Dixon equations for the spin tensor alongside the rigid-body tidal tensor and the first-order post-Newtonian orbital corrections demonstrates that both the covariant transport and the Newtonian torque induce a step-function evolution in the projected rotational velocity, $v \sin i$.

This localization enables a clean parametric separation between the two driving torques. The relativistic geodetic contribution to the velocity shift scales linearly with the equatorial rotation speed $v_{\rm rot}$. In contrast, the shift induced by the classical quadrupole torque is independent of $v_{\rm rot}$ and scales only with the stellar ellipticity $q$ and radius. The Newtonian tidal torque therefore acts as a rotation-independent baseline for oblate configurations, while the geodetic signal grows monotonically with rotation speed. This opposite dependence on $v_{\rm rot}$ constitutes a key discriminative handle.

To quantify the expected signal, we evaluate an isotropic distribution of initial rotational axes and viewing geometries over a 40-year period across a parameter grid $v_{\rm rot} \in [200, 400]\,\kms$ and $q \in [0.05, 0.3]$. The statistical distribution of the absolute accumulated shift $|\Delta v \sin i|$ leads to median values between $3\,\kms$ and $6.3\,\kms$ for typical configurations. Edge-on geometries and favourable pole orientations generate maximal absolute shifts reaching $46.1\,\kms$ for the fastest and most oblate models. Over multiple periapsis passages, the precessional motion causes the projcted velocity to oscillate within the geometric bounds imposed by the intrinsic rotation speed, rather than accumulating monotonically.

The post-Newtonian advance of the orbital ellipse systematically alters the viewing geometry of the orbit between successive revolutions. This modulation of the periapsis-step amplitude as a function of epoch provides a time-domain discriminant. Tracking the line broadening across several orbital periods thus allows one to disentangle the relativistic geodetic precession from the classical quadrupole torque, isolating the covariant component. We conclude that S301 constitutes a target for current and next-generation infrared spectrographs to test the Schwarzschild metric around Sgr~A$^\ast$. While orbital astrometry traces geodesic motion, measuring the spin precession directly tests Fermi-Walker transport and the spin-curvature coupling. This isolates the spatial curvature of the spacetime, providing a novel constraint on possible deviations from general relativity that complements existing tests based on orbital astrometry and redshift measurements.

\begin{acknowledgments}
We acknowledge support by the National Foreign Expert Program (H). DC and JC acknowledge the financial support from ANID-FONDECYT Regular 1251444. The research of DC has been funded by the Alexander von Humboldt Foundation.

\end{acknowledgments}

\section*{End Matter}
\appendix
\section{Detailed Derivation of Equation (6)}

To calculate the classical rigid-body torque, we evaluate the spatial gradient of the central potential $\Phi = -M/r$. The local tidal tensor $\mathcal{E}$ has components

\begin{equation}
\label{eq:tidal_tensor}
\mathcal{E}_{ij} = -\partial_i \partial_j \Phi = \frac{M}{r^3} (3n_i n_j - \delta_{ij}).
\end{equation}

\noindent
The interaction energy $U$ between the tidal field and the stellar mass distribution depends on the moment of inertia tensor $I_{ij}$, expressed as

\begin{equation}
\label{eq:interaction_energy}
U = -\frac{1}{2} \mathcal{E}_{ij} I_{ij}.
\end{equation}

\noindent
We model the star as an axisymmetric body with the symmetry axis aligned with the unit rotation vector $\bm{s}$. The moment of inertia tensor takes the form

\begin{equation}
\label{eq:inertia_tensor}
I_{ij} = I_1 \delta_{ij} + (I_3 - I_1) s_i s_j,
\end{equation}

\noindent
where $I_3$ is the principal moment of inertia along the rotational axis, and $I_1$ is the transverse moment. Substituting Eq.~(\ref{eq:inertia_tensor}) into Eq.~(\ref{eq:interaction_energy}) and applying the trace-free property of the tidal tensor produces the effective interaction energy

\begin{equation}
\label{eq:interaction_energy_final}
U = -\frac{3M}{2r^3} (I_3 - I_1) (\bm{n} \cdot \bm{s})^2.
\end{equation}

\noindent
The resulting torque $\bm{N}_{\text{quad}} = - \bm{s} \times \nabla_{\bm{s}} U$ acts perpendicular to the rotation vector, taking the form

\begin{equation}
\label{eq:torque_derivation}
\bm{N}_{\text{quad}} = -\frac{3M}{r^3} (I_3 - I_1) (\bm{n} \cdot \bm{s}) (\bm{n} \times \bm{s}).
\end{equation}

\end{document}